\documentclass[12pt]{article}
	\pdfoutput=1	
\usepackage{amsmath,amssymb}
\usepackage{graphicx}
\usepackage[pdftex]{hyperref}
\usepackage{caption}
\DeclareGraphicsExtensions{.eps,.bmp,.wmf,.jpg}
\numberwithin{equation}{section}
\def\be{\begin{equation}}
\def\ee{\end{equation}}

\def\bea{\begin{eqnarray}}
\def\eea{\end{eqnarray}}

\title{Reconstructing modified gravity with holographic vacuum energy density}
\author{L.N. Granda\thanks{luis.granda@correounivalle.edu.co}\\{\it Departamento de Fisica, Universidad del Valle}\\{\it A.A. 25360, Cali, Colombia}}
\date{}
\begin{document}
\maketitle

\begin{abstract}
A reconstruction of modified gravity is proposed by establishing a correspondence between the effective density of the modified gravity and the holographic density. The non-homogeneous term in the modified Friedmann equation, generated by the vacuum (holographic) energy density, lead to reconstructed models that contain explicitly, as part of the solution, the Einstein-Hilbert term. It was shown that the $\Lambda$CDM-type cosmic histories allow the general relativity Lagrangian with cosmological constant as a particular reconstructed solution. The Starobinsky term appears in some reconstructed solutions, and an approximate reconstruction of the Chaplygin gas cosmology was performed in terms of elementary functions of curvature.
\begin{description}
\item[PACS numbers]
98.80.-k, 95.36.+x, 04.50.Kd
\end{description}
\end{abstract}

\maketitle


\section{\label{intro}Introduction}
Among the alternatives to the explanation of dark energy (for review see \cite{copeland06, sahnii04, padmanabhan, sergeiod}), a very promising approach to dark energy is related with the modified theories of gravity known as f(R) gravity (see \cite{sodintsov1, sotiriou, tsujikawa0, sodintsov1a, nojirioo, sinodin} for reviews). In these models the large-distance behavior of gravity emerges from the modification of geometrical terms in the gravitational action. Modified gravity gives a natural unification of the early time inflation and late-time acceleration due  to different role of gravitational terms relevant at small and at large curvature and may naturally describe the transition from deceleration to acceleration in the cosmological dynamics. 
The $f(R)$ theories have been intensively studied to explain the late time accelerated expansion and many types of modifications to the Einstein-Hilbert (EH) action have been proposed so far \cite{capozziello, capozziello1, sodintsov, carroll, faraoni, dobado, anthoni, barrow1, nojiri,  elizalde, troisi, allemandi, koivisto, brevik, sodintsov2, nojiri09, carloni, dunsby, nojiri1, nojiri2, olmo, hu}. Particularly, in \cite{sodintsov2} a general scheme for reconstructing from viable FRW cosmologies was proposed, where the obtained modified gravity models were consistent with solar system tests and describe different cosmological phases going from matter dominance to current accelerated expansion consistent with observations. In \cite{nojiri09}, a general cosmological reconstruction method in terms of the e-folding variable was developed and reconstruction of modified gravity with an extra scalar field was considered. This method allows to reproduce viable $f(R)$ gravities that pass local tests and unify early-time inflation with late-time acceleration. In \cite{carloni} a new way of reconstructing $f(R)$ is proposed, by imposing dynamical restrictions on the cosmic parameters like scale factor, Hubble parameter, decelerating parameter and higher order parameters. In \cite{dunsby}, it was shown that the only $f(R)$ model that allows an exact  $\Lambda$CDM expansion is the standard EH Lagrangian with a positive cosmological constant, and for more general functions of $R$, the only way to reproduce $\Lambda$CDM is by using additional degrees of freedom added to the matter sector. In \cite{sinodin1}, the extension of reconstructing scheme is applied to scalar-tensor, Gauss-Bonnet modified gravity and scalar-Gauss-Bonnet gravity, showing that the phases of mater dominance followed by the dark energy dominance may emerge also from these theories.
The $f(R)$ theories may also be relevant to early-time inflation at large $R$, due to its non-linear character \cite{starobinsky}. The unified description of early time inflation and late time accelerated expansion were also considered in \cite{nojiri5, nojiri6, nojiri7, nojiri8, odinoik}. In \cite{nojiri5} positive powers of curvature have been introduced to account for the inflationary epoch and negative powers of curvature to explain late time acceleration. A model behaving as a positive power of the curvature at large curvature has been proposed in \cite{nojiri7}, in order to unify inflation with late-time $\Lambda$CDM epoch. In \cite{nojiri8}, a step-class models of modified gravity have been considered, showing that the early inflation and late time accelerated expansion arise in these models in a unified way.  
Any realistic model of modified gravity should pass not only the local tests where the average density of matter is high compared with that of the universe, but also the observational cosmological restrictions. To pass solar system tests the model has to implement the so called chameleon mechanism \cite{hu, tsujikawa, brax} which gives a large enough mass to the scalar field to avoid measurable corrections to the local gravity  phenomena which is well described by the general relativity. Models that can satisfy both cosmological and local gravity constraints have been proposed in \cite{hu, astarobinsky, appleby1, sergeid1, sergeid2, eelizalde}. Exact solutions explaining the current accelerated expansion are presented in \cite{bamba1, barrow, clifton, capozz, capozz1, capozz2}.\\
Unfortunately the complexity of the fourth-order field equations of the $f(R)$ theories make them difficult to integrate and to find analytical (and numerical) solutions, which additionaly, should obey a number of restrictions and be consistent  with observations \cite{amendola1,amendola2}. 
However an interesting approach to solve this problem is to use the technique of reconstruction where one assumes a given cosmological expansion law and use the inverse method in the field equations to find the $f(R)$ function or class of functions that give rise to the proposed solution. One important issue in this method is the condition that the scalar curvature $R$ should be analytically invertible (with respect to the main variable of the proposed solution), allowing to express the the main evolutionary magnitudes like $H$ and $\dot{H}$ in terms of $R$, turning at the end the equation of motion into a differential equation in the $R$ space. This reconstruction scheme has severe limitations because the proposed solutions should be simple enough to permit the inverse integration of the equations to obtain the analytic expression for $f(R)$. Besides that, the reconstructed model is usually very constrained and is difficult to obtain, starting from the reconstructed Lagrangians which are often given by a very complicated expressions, other solutions different form the one used to its reconstruction. 
A typical and relevant example is the power-law solution in the FRW background, since these solutions represent important asymptotical states in the cosmological evolution, when the matter content is dominated by a certain type of matter (radiation, cold dark matter, dark energy, etc.). These solutions allow the integration in the $R$ space, giving $f(R)\propto R^n$ \cite{goheer}, which is quite strong constraint.
On the other hand, the dark energy that is well described by the cosmological constant, which in the frame of the quantum field theory (QFT) is the vacuum energy, faces the known problem of the fine tuning. The vacuum energy is described within the framework of the QFT in Minkowski background. At cosmological scales, when the effect of gravity is relevant, the above description of vacuum energy fails and is very likely that the correct value of the vacuum energy would be predicted by a fundamental theory of quantum gravity (QG). While the lack of QG theory keeps us away from the definite solution of the vacuum energy problem, the holographic principle \cite{thooft, susskind}, which incorporates some features of QG, represents an interesting approach to the dark energy (cosmological constant) problem, where this principle establishes an infrared cut-off for the so called holographic energy density related with cosmological scales \cite{cohen, hsu, li, granda1}. Viewing the modified f(R) gravity models as an effective description of the underlying theory of dark energy, it is interesting to study how the f(R) gravity can be mapped to the holographic energy density as an effective theory. \\
In the present paper we study the correspondence between the effective energy density of the $f(R)$ gravity and the holographic energy density as proposed in \cite{granda1} and its implications in cosmological scenarios with accelerated expansion.  An important result is that the Einstein-Hilbert term appears naturally in all the reconstructed $f(R)$ models, due to the source term given by the holographic density. 
This paper is organized as follows. In section II we present the equations of motion in general and in the FRW background and introduce the reconstruction method. In section III we apply the reconstruction technique to some cosmological solutions. In section IV we present a discussion.
\section{Field Equations and reconstruction scheme}
The action for the $f(R)$-gravity with the source of matter id given by 
\be\label{eq1}
S=\int d^4x\sqrt{-g}\left[\frac{1}{2\kappa^2}f(R)+{\cal L}_m(\psi)\right]
\ee
where $\kappa^2=8\pi G$ and ${\cal L}_m$ is the Lagrangian density for the matter component, which can represent the usual usual baryonic matter, dark matter as well as more exotic type of matter that we will associate with the vacuum energy provided by the holographic principle. Variation with respect to the metric gives the following equation of motion
\be\label{eq2}
f'(R)R_{\mu\nu}-\frac{1}{2}g_{\mu\nu}f(R)+\left(g_{\mu\nu}\Box-\nabla_{\mu}\nabla_{\nu}\right)f'(R)=
\kappa^2 \left(T_{\mu\nu}^{(m)}+T_{\mu\nu}^{(hol)}\right)
\ee
where $T^{(m)}_{\mu\nu}$ is the matter energy-momentum tensor, $T_{\mu\nu}^{(hol)}$ refers to the holographic energy and the prime indicates the derivative with respect to $R$. Taking the trace of eq. (\ref{eq2})  leads to 
\be\label{eq3}
Rf'(R)-2f(R)+3\Box f'(R)=\kappa^2 T
\ee
where $T$ is the trace of the energy-momentum tensor. Assuming the flat FRW background, the time and spatial components of the Eq. (\ref{eq2}) take the form
\begin{align}\label{eq2a}
3H^2f'(R)-\frac{1}{2}\left(Rf'(R)-f(R)\right)+3H\dot{R}f''(R)=	\kappa^2\left(\rho_m+\rho_{\Lambda}\right)
\end{align}
and
\be\label{eq2b}
\left(3H^2+\dot{H}\right)f'(R)-\frac{1}{2}f(R)-2H\dot{R}f''(R)-f''(R)\ddot{R}-f'''(R)\dot{R}^2=\kappa^2\left(p_m+p_{\Lambda}\right)
\ee
where $\rho_m$ and $p_m$ are the density and pressure due to the matter content. All curvature dependent terms in the l.h.s. of both equations give the generalization to the EH model, and these equations reduce to the known cosmological equations of general relativity by setting $f(R)=R$. 
One can also separate the time and spatial components of the Einstein's tensor from equations (\ref{eq2a}) and (\ref{eq2b}) respectively to write the equations in the standard form, where the additional $f(R)$-dependent terms can be interpreted as effective density and pressure. In this case the equations (\ref{eq2a}) and (\ref{eq2b}) take the form
\be\label{eq3a}
H^2=\frac{\kappa^2}{3}\left(\rho_f+\tilde{\rho}_m+\tilde{\rho}_{\Lambda}\right)
\ee
with the effective "curvature" density given by
\be\nonumber
\rho_f= \frac{1}{\kappa^2f'(R)}\left(\frac{1}{2}\left(Rf'(R)-f(R)\right)-3H\dot{R}f''(R)\right)
\ee
and
\be\nonumber
\tilde{\rho}_m=\frac{\rho_m}{f'(R)},\;\;\; \tilde{\rho}_{\Lambda}=\frac{\rho_{\Lambda}}{f'(R)}
\ee
\be\label{eq3b}
-\left(3H^2+2\dot{H}\right)=\kappa^2\left(p_f+\tilde{p}_m+\tilde{p}_{\Lambda}\right)
\ee
with the effective "curvature" pressure given by
\be\nonumber
p_f=\frac{1}{\kappa^2f'(R)}\Big(\frac{1}{2}\left(f(R)-Rf'(R)\right)+2H\dot{R}f''(R)+\ddot{R}f''(R)+\dot{R}^2f'''(R)\Big)
\ee 
and 
\be\nonumber
\tilde{p}_m=\frac{p_m}{f'(R)},\;\;\; \tilde{p}_{\Lambda}= \tilde{p}_{\Lambda}=\frac{p_{\Lambda}}{f'(R)}.
\ee
The equation (\ref{eq3}) has also an interesting interpretation if one introduces the so called "scalaron" field $f'(R)$.  By writing the trace equation (\ref{eq3})  in the form
\be\label{eq5} 
\Box f'(R)=\frac{\partial V_{eff}}{\partial f'(R)}
\ee
where 
\be\label{eq6}
\frac{\partial V_{eff}}{\partial f'(R)}=\frac{1}{3}\left(2f(R)-Rf'(R)+\kappa^2T\right)
\ee
We may interpret the model (\ref{eq1}) in the Einstein frame (which is specially useful to study the inflation) by performing a conformal transformation of the metric with the function $f'(R)$ \cite{barrow2}
\be\label{eq13}
g_{\mu\nu}\rightarrow \tilde{g}_{\mu\nu}=f'(R)g_{\mu\nu}=e^{-\sqrt{\frac{2}{3}}\kappa\phi}g_{\mu\nu},
\ee
the action takes the form
\begin{align}\label{eq14}
S=\int d^4x\sqrt{-\tilde{g}}
\left[\frac{1}{2\kappa^2}\tilde{R}-\frac{1}{2}\tilde{g}^{\mu\nu}\partial_{\mu}\phi\partial_{\nu}\phi-V(\phi)+\right.
\left. {\cal L}_m(e^{\sqrt{\frac{2}{3}}\kappa\phi}\tilde{g}_{\mu\nu},\psi)
\right]
\end{align}
with the potential
\be\label{eq15}
V(\phi)=\frac{R(\phi)f'(R(\phi))-f(R(\phi))}{2\kappa^2 f'(R(\phi))^2}
\ee
in which the new scalar field "scalaron" couples minimally to the scalar curvature but becomes coupled to the matter sector. \\
Let us turn to the matter content of the model in the FRW background. To the energy-momentum tensor the main contribution is given by the matter content (including baryonic and dark matter) and the vacuum energy, but for the late-time cosmological scenarios we will consider that the dominant contribution is given by the vacuum energy described by the holographic density \cite{granda1, granda2}, given by
\be\label{eq16}
\rho_{\Lambda}=\frac{3}{\kappa^2}\left(\alpha H^2+\beta \dot{H}\right)
\ee
which behaves a a perfect fluid and obeys the conservation equation
\be\label{eq17}
\dot{\rho}_{\Lambda}+3H\left(\rho_{\Lambda}+p_{\Lambda}\right)=0
\ee
This model can be reproduced from the generalized HDE model introduced in \cite{nojiodin1}, where different infrared cutoffs were proposed, that explain a variety of late-time cosmological scenarios, including solutions with crossing of the phantom divide. It should be noted that the equation (\ref{eq2b}) is not independent of the equation (\ref{eq2a}) since the equation  (\ref{eq2b}) can be obtained by linearly combining the equation (\ref{eq2a}) with its time derivative  and using the continuity equation for the matter component. Hence, any solution of the modified Friedmann equation  (\ref{eq2a}) automatically solves the modified equation (\ref{eq2b}) for the pressure. Thus, to perform the reconstruction  it is sufficient to solve the Friedmann equation.
Then, for the reconstruction we use the modified Friedmann equation  (\ref{eq16})  with the source term given by the holographic density $\rho_{\Lambda}$, as follows
\be\label{eq18}
3H^2f'(R)-\frac{1}{2}\Big(Rf'(R)-f(R)\Big)+3H\dot{R}f''(R)=3\left(\alpha H^2+\beta\dot{H}\right)
\ee
The reconstruction technique assumes that the expansion history of the universe is known, and by this solution we re-express the coefficients of the equation (\ref{eq2a}) in terms of the scalar curvature, which turns the modified Friedmann equation into a differential equation in the $R$-space. The explicit reconstruction is possible only in the case when the expression for Ricci scalar $R$ can be solved algebraically with respect to one of the variables, $t$, $a$, $\ln a$, etc. Thus, for instance, if the time variable can be expressed as $t=g(R)$, which allows to write functions like $a(t)$, $H(t)$, $\dot{H}$ in terms of $R$. The method is interesting whenever one can reconstruct $f(R)$ for relevant and viable cosmological solutions. \\
Starting from the modified Friedmann equation  (\ref{eq18}) and by known evolutionary history of the universe, we try to reconstruct the $f(R)$ model assuming that the vacuum energy is dominant and is described by the holographic principle. Following the lines of reconstruction as proposed in \cite{nojiri09, carloni, dunsby}, and using the e-folding variable $x=\ln a$, we can write the Ricci scalar as (using $\frac{d}{dt}=H\frac{d}{dx}$)
\be\label{eq19}
R=6\left(2H^2+\dot{H}\right)=12H^2+3\frac{dH^2}{dx}
\ee
and the Eq. (\ref{eq18}) takes the form
\be\label{eq20}
3H^2f'(R)-\frac{1}{2}\Big(Rf'(R)-f(R)\Big)+3H^2\frac{dR}{dx}f''(R)=3\left(\alpha H^2+\frac{1}{2}\beta\frac{dH^2}{dx}\right)
\ee
If a cosmological solution is given as a function of $x$ and one can write the Hubble parameter as 
\be\label{eq20}
H^2=\Phi(x),
\ee
then the scalar curvature takes the form
\be\label{eq21}
R=12\Phi(x)+3\frac{d\Phi(x)}{dx}.
\ee
In the cases when this equation can be solved with respect to $x$ as $x=x(R)$, which allows to write the function $\Phi(x)$ as $\Phi(x(R))\equiv\Phi(R)$ and the derivative with respect to $x$ as
\be\label{eq22}
\frac{d}{dx}=\left(\frac{dx}{dR}\right)^{-1}\frac{d}{dR},
\ee
at the end one can write the modified Friedmann equation (\ref{eq20}) as follows
\be\label{eq23}
\begin{aligned}
&3\Phi(R)\left(\frac{dx}{dR}\right)^{-1}f''(R)+\left(3\Phi(R)-\frac{1}{2}R\right)f'(R)+\frac{1}{2}f(R)
=\\& 3\left(\alpha \Phi(R)+\frac{1}{2}\beta\frac{d\Phi(R)}{dR}\left(\frac{dx}{dR}\right)^{-1}\right),
\end{aligned}
\ee 
which in compact form can be written as
\be\label{eq24}
C_1(R)f''(R)+C_2(R)f'(R)+\frac{1}{2}f(R)=C(R)
\ee
with coefficients $C_1$, $C_2$ and free term $C$
\be\label{eq25}
\begin{aligned}
&C_1(R)=3\Phi(R)\left(\frac{dx}{dR}\right)^{-1},\;\;\;
C_2(R)=3\Phi(R)-\frac{1}{2}R\\
&C(R)=3\left(\alpha \Phi(R)+\frac{1}{2}\beta\frac{d\Phi(R)}{dR}\left(\frac{dx}{dR}\right)^{-1}\right)
\end{aligned}
\ee
which is a non-homogeneous second order differential equation in the $R$-space. Note that the non-homogeneous term comes from the vacuum energy, which in the present case is described by the holographic density. So the success of the reconstruction depends how adequate the coefficients $C_1$ and $C_2$ are to be able to integrate the equation (\ref{eq24}). The coefficients $C_1$ and $C_2$ should correspond to a viable and consistent with observations expansion history. The above reconstruction procedure can also be performed with respect to any other variable, $y$ for instance, which could represent the time $t$, slow-roll $x=\ln a$, scale factor $a$, etc.. In these cases one should be able to solve explicitly the $y$-variable in terms of the curvature, and them write and solve the equation (\ref{eq18}) in the corresponding  $y$-space. An especially simple form acquires the coefficient $C(R)$ for the Ricci density \cite{gao} that takes place for $\alpha=2\beta$, giving $C(R)=\frac{1}{2}\beta R$. Thus, in the solutions considered bellow, the Ricci limit can be obtained by setting $\alpha=2\beta$. 
\section{Reconstructing models}
{\it Power-law expansion}\\
As a first case we consider the power-law expansion described as follows
\be\label{eq26}
H=H_0 a^{-\frac{1}{p}}=H_0 e^{-\frac{1}{p}x}
\ee
which gives the known solution
\be\label{eq27}
a=a_0 t^p,\;\;\;\; H=\frac{p}{t}
\ee
It is easy to check that for this solution one finds
\be\label{eq28}
\Phi(R)=\frac{p}{6(2p-1)}R,\;\;\; \left(\frac{dx}{dR}\right)^{-1}=-\frac{2}{p}R
\ee
which gives for $C_1$, $C_2$ and $C$:
\be\label{eq29}
C_1=-\frac{1}{2p-1}R^2,\;\;\; C_2=\frac{1-p}{2(2p-1)}R,\;\;\;
C=\frac{\alpha p-\beta}{2(2p-1)}R.
\ee
Substituting these coefficients in the equation (\ref{eq24}), one finds the following general solution
\be\label{eq30}
f(R)=\frac{\alpha p-\beta}{p}R+b_1 R^{p_1}+b_2 R^{p_2}
\ee
where
\be\nonumber
p_1=3-p-\sqrt{p^2+10p+1},\;\; p_2=3-p+\sqrt{p^2+10p+1}
\ee
and $b_1, b_2$ are the constants of integration. Where $p>1/2$ generates both, negative powers of $R$ ($p_1<0$) and positive powers of $R$ ($p_2>0$). Note that we can normalize the expression (\ref{eq30}) by taking and dropping the overall factor $\frac{\alpha p-\beta}{p}$ and rescaling the integration constants by the inverse of this factor. In this case, the first term gives the usual EH Lagrangian, while the other terms coming from the homogeneous solution, give the correction due to the more general nature of the modified Friedmann equation. One can also set the coefficient $\frac{\alpha p-\beta}{p}=1$ in order to satisfy the EH limit, in which case and setting $b_1=b_2=0$ one obtains the results of the model \cite{granda1, granda2}. With $$p=\beta/(\alpha-1),$$ the condition ($p>0,p_1>0,p_2>0$) reduces to $\beta<0$ and $\alpha<1+2\beta$ or $\beta>0$ and $\alpha>1+2\beta$, and the condition ($p>0, p_1<0, p_2>0$) reduces to $\beta<0$ and $1+2\beta<\alpha<1$ or $\beta>0$ and $1<\alpha<1+2\beta$. In the present solution, the linear in $R$ term is possible thanks to the vacuum contribution $\rho_{\Lambda}$ coming from the proposed holographic density.\\
The phantom power-law expansion is described by
\be\label{eq31}
a=a_0\left(\frac{t_0}{t_c-t}\right)^p,\;\;\; H=\frac{p}{t_c-t}.
\ee
From $x=\ln a$ follows that $\frac{t_0}{t_c-t}=e^{x/p}$, giving 
\be\label{eq32}
\Phi(R)=\frac{p}{6(2p+1)}R,\;\;\; \left(\frac{dx}{dR}\right)^{-1}=\frac{2}{p}R.
\ee
for the coefficients $C_1$, $C_2$ and $C$ it is obtained 
\be\label{eq29}
C_1=\frac{1}{2p+1}R^2,\;\;\; C_2=-\frac{p+1}{2(2p+1)}R,\;\;\;
C=\frac{\alpha p+\beta}{2(2p+1)}R.
\ee
these coefficients can be obtained from (\ref{eq29}) by changing $p\rightarrow -p$, and so the general solution of the equation (\ref{eq24}) is
\be\label{eq30}
f(R)=\frac{\alpha p+\beta}{p}R+b_1 R^{p_1}+b_2 R^{p_2}
\ee
with
\be\nonumber
p_1=3+p-\sqrt{p^2-10p+1},\;\; p_2=3+p+\sqrt{p^2-10p+1}
\ee
where $p\ge5+2\sqrt{6}$.
Similar solution, without the linear term, can be obtained for power-law expansion without introducing matter term \cite{nojiri09}. The EH limit is satisfied by setting $p=\beta/(1-\alpha)$. In this case, the condition ($p>0,p_1>0, p_2>0$) reduces to $\beta<0$ and $\alpha>1-2\beta$ or $\beta>0$ and $
\alpha<1-2\beta$ and the condition ($p>0,p_1<0, p_2>0$) reduces to $\beta<0$ and $1<\alpha<1-2\beta$ or $\beta>0$ and $1-2\beta<\alpha<1$. \\
{\it Little Rip}\\
As the second case we consider the following expansion law, where $\Lambda$ is a positive constant
\be\label{eq31}
H^2=h_0^2 x+\Lambda.
\ee
The effective equation of state (EoS) derived from this expansion law is given by
\be\label{eq31a}
w=-1-\frac{1}{3H^2}\frac{dH^2}{dx}=-1-\frac{h_0^2}{3(h_0^2x+\Lambda)},
\ee 
which describes current ($x=0$) phantom expansion with $w_0=-1-h_0^2/(3\Lambda)$. If one fixes the current EoS, for instance $w=-1.05$, then we can find the relation between $h_0^2$ and $\Lambda$ as $h_0^2/\Lambda=0.15$. In this scenario, $w$ asymptotically approaches the de Sitter $w=-1$ at $x\rightarrow \pm\infty$. From (\ref{eq31}) it follows that $\dot{H}=h_0^2/2$, and hence  $H=\frac{1}{2}h_0^2\left(t-t_0\right)$, which gives
\be\label{eq31b}
a(t)=a_0 e^{\frac{1}{4}h_0^2(t-t_0)^2}
\ee
since $H$increases with time as $H=\frac{1}{2}h_0^2\left(t-t_0\right)$, but remains finite at finite time, then it describes a Little Rip solution. This time dependence for $H$ was also considered in \cite{nojiri09}, where the reconstructed $f(R)$ was expressed trough the Kummers series. 
From (\ref{eq20}) and (\ref{eq21}) it follows that 
\be\label{eq32}
\Phi(R)=\frac{R-3h_0^2}{12},\;\;\; x=\frac{R-3(h_0^2+4\Lambda)}{12h_0^2},
\ee
leading to the coefficients
\be\label{eq33}
C_1=3h_0^2\left(R-3h_0^2\right),\;\;\; C_2=-\frac{1}{4}\left(R+3h_0^2\right),\;\;\;
C=\frac{1}{4}\left(\alpha R-3\alpha h_0^2+6\beta h_0^2\right)
\ee
Solving the modified Friedmann equation (\ref{eq24}), in general, the solution is obtained through the Laguerre polynomials and GammaRegularized functions, but one can propose a simple particular solution to this equation given by 
\be\label{eq34}
f(R)=\alpha_0+\alpha_1 R+\alpha_2 R^2.
\ee
By replacing this solution in (\ref{eq24}), the following relations are found 
$$ \alpha_0=3h_0^2\left(3\alpha_2 h_0^2+\beta\right), \;\;\; \alpha_1=\alpha-18\alpha_2 h_0^2$$
where the constant $\alpha_2$ is arbitrary. By setting $\alpha_1=1$ (to get the EH term), one can fix $\alpha_0$ and $\alpha_2$ in terms of $h_0$, $\alpha$ and $\beta$
\be\label{eq35}
f(R)=R+\frac{\alpha-1}{18h_0^2}R^2+\frac{1}{2}\left(\alpha+6\beta-1\right)h_0^2
\ee
This model contains the Starobinsky $R^2$ term (for $\alpha>1$) which gives a singularity-free cosmology, and the inflation based on this model remains in good agreement with the current measurements of the cosmic microwave background. \\
{\it $\Lambda$CDM-type}\\
For the following case we consider the solution
\be\label{eq36}
H^2=h^2 e^{-\lambda x}+\Lambda,
\ee
which gives the effective EoS
\be\label{eq36a}
w=-1+\frac{1}{3}\frac{\lambda h^2 e^{-\lambda x}}{h^2 e^{-\lambda x}+\Lambda}
\ee
which describes quintessence or phantom evolution, depending on the sign of $\lambda$. Once one defines the expansion law encoded in $\lambda$, then the current EoS $w_0$ becomes defined by the relation $\Lambda/h^2$. From (\ref{eq20}) and (\ref{eq21}) it follows 
\be\label{eq37}
\Phi=\frac{R}{3(4-\lambda)}-\frac{\lambda\Lambda}{4-\lambda},\;\;\; x=-\frac{1}{\lambda}\ln \frac{R-12\Lambda}{3h^2(4-\lambda)}
\ee
leading to the following coefficients
\be\label{eq38}
\begin{aligned}
&C_1=\frac{\lambda(R-3\lambda\Lambda)(R-12\Lambda)}{\lambda-4},\;\;\; C_2=\frac{(\lambda-2)R-6\lambda\Lambda}{2(4-\lambda)},\\
&C=\frac{1}{2(4-\lambda)}\left[\left(2\alpha-\beta\lambda\right)R+6\lambda\left(2\beta-\alpha\right)\Lambda\right]
\end{aligned}
\ee
with these coefficients, the modified Friedmann equation (\ref{eq24}) have a very simple particular solution given by
\be\label{eq39}
f(R)=\frac{1}{2}\left(2\alpha-\beta\lambda\right)R+3\beta\lambda\Lambda.
\ee
This solution is possible thanks to the non-homogeneous term given by the holographic energy density, since it depends on $\alpha$ and $\beta$ and disappears if $\alpha=\beta=0$ (there is not elementary polynomial solution in this case). So in the framework of the holographic dark energy (HDE), the reconstructed $f(R)$ gravity from the $\Lambda$CDM-type cosmological solution, exactly leads to the EH Lagrangian with cosmological constant, without introducing any matter source (the $\Lambda$CDM cosmology corresponds to $\lambda=3$). In \cite{dunsby} it was shown that it is not possible to mimic the $\Lambda$CDM expansion, for a vacuum universe, with a real valued $f(R)$, but if a source term is introduced, then there were found cases where the reconstructed model could obey the $\Lambda$CDM expansion history. Thus, for a universe filled with dust-like matter, it was shown in \cite{dunsby} that the only real valued $f(R)$ that reproduces the $\Lambda$CDM expansion is the Einstein-Hilbert Lagrangian with positive cosmological constant. The presence of the cosmological constant in the reconstructed solution was important to obtain the result $f(R)=R-2\Lambda$, since if $\Lambda=0$ is assumed, then a family of $f(R)$ solutions appears \cite{dunsby} that can reproduce a dust-like expansion without a cosmological constant. A $\Lambda\ne 0$ in the reconstructed history breaks de degeneracy in $f(R)$ models and leads to the unique solution. If along with the dust-like matter, another matter content is considered, like a stiff fluid with equation of state $w=1$, or non isentropic perfect fluid \cite{dunsby}, then the reconstructed $f(R)$ theory contains the Einstein-Hilbert Lagrangian with cosmological constant plus higher powers of the Ricci scalar (that are irrelevant for late-time universe), leading also to practically exact $\Lambda$CDM expansion history. In the present work, the reconstructed $f(R)$ (\ref{eq35}) from the little Rip solution (\ref{eq31}) can also reproduce the $\Lambda$CDM expansion (neglecting the $R^2$ term at late times), and if we take $\alpha=1$ then the $R^2$ term disappears, leading to the EH Lagrangian with a cosmological constant given by $-3\beta h_0^2/2$. So, in the framework of the HDE, the standard EH term with cosmological constant can reproduce little rip solutions as described by (\ref{eq31}).\\
There is another possible polynomial solution if we fix the parameter $\lambda$. One can propose the solution
\be\label{eq40}
f(R)=\alpha_0+\alpha_1 R+\alpha_2 R^2+\alpha_3 R^3
\ee
after replacing this solution into the Friedmann equation (\ref{eq24}) with coefficients given by (\ref{eq38}), we can see that the term with $R^3$ satisfies the solution for $\lambda=-1/5$, leaving arbitrary the coefficient $\alpha_3$ and
\be\label{eq41}
\begin{aligned}
\alpha_0=-\frac{3}{125}\left(25\beta\Lambda-576\alpha_3\Lambda^3\right)&,\;\;
\alpha_1=\frac{1}{250}\left(250\alpha+25\beta+35424\alpha_3\Lambda^2\right),\\ & \alpha_2=\frac{198\alpha_3\Lambda}{5}
\end{aligned}
\ee
the cosmological history that originates this model corresponds to the solution
\be\label{eq42}
H^2=h^2 e^{x/5}+\Lambda
\ee
which describes phantom expansion according to (\ref{eq36a}). \\
{\it Chaplygin gas}\\
Among the different models proposed to describe the observed accelerated expansion of the universe, the Chaplygin gas \cite{kamenshchik} has attracted much attention as it interpolates between the dust-matter dominated phase at early times and an universe dominated by the cosmological constant at late times, giving a unified description of dark matter and dark energy with an equation of state in the range $-1\le w\le 0$. 
The Chaplygin gas solution can be written as  
\be\label{eq43}
H^2=\left(A+B e^{-6x}\right)^{1/2}
\ee
where $A$ and $B$ are positive constants. 
To find an $f(R)$ model that reproduces the Chaplygin gas cosmology  (\ref{eq43}) , we first introduce the new variable $y$ as follows
\be\label{eq44}
y=H^2=\left(A+B e^{-6x}\right)^{1/2}
\ee
Then,
$$B e^{-6x}=y^2-A,\;\; \dot{H}=\frac{1}{2}\frac{dH^2}{dx}=\frac{1}{2}\frac{dy}{dx}=-\frac{3}{2y}\left(y^2-A\right),$$ giving
\be\label{eq45}
R=\frac{3}{y}\left(y^2+3A\right).
\ee
Next, we writhe the modified Friedmann equation (\ref{eq20}) in terms of $y$ and using $\frac{d}{dx}=\left(\frac{dx}{dy}\right)^{-1}\frac{d}{dy}$ as follows
\be\label{eq46}
\begin{aligned}
&3H^2\left(\frac{dR}{dy}\right)^{-1}\left(\frac{dx}{dy}\right)^{-1}\frac{d^2f}{dy^2}+\Big[\left(3H^2-\frac{1}{2}R\right)\left(\frac{dR}{dy}\right)^{-1}-\\& 3H^2\left(\frac{dR}{dy}\right)^{-2}\frac{d^2R}{dy^2}\left(\frac{dx}{dy}\right)^{-1}\Big]\frac{df}{dy}
+\frac{1}{2}f(y)=3\left(\alpha H^2+\frac{1}{2}\beta \frac{dH^2}{dy}\left(\frac{dx}{dy}\right)^{-1}\right)
\end{aligned}
\ee
using (\ref{eq44})-(\ref{eq45}) we finally arrive at the following non-homogeneous differential equation in the y-space
\be\label{eq47}
\begin{aligned}
&-6y^3\left(y^4-4Ay^2+3A^2\right)\frac{d^2f(y)}{dy^2}+y^2\left(y^4-12Ay^2-9A^2\right)\frac{df(y)}{dy}+\\ &y\left(y^2-3A\right)^2f(y)=3\left(y^2-3A\right)^2\left[\left(2\alpha-3\beta\right)y^2+3\beta A\right]
\end{aligned}
\ee
It is not possible to find a general analytical solution to this equation, but there are particular cases and an approximation that could be acceptable for late time universe. For the simple case of $A=0$ corresponding to usual dark matter dominance with $\rho\propto a^{-3}$, the Eq. (\ref{eq47}) reduces to
\be\label{eq48}
-6y^3\frac{d^2f(y)}{dy^2}+y^2\frac{df(y)}{dy}+yf(y)=3\left[\left(2\alpha-3\beta\right)y^2+3\beta A\right]
\ee
which gives the particular solution
\be\label{eq49}
f(y)=\frac{3}{2}\left(2\alpha-3\beta\right)y,
\ee
then. according to (\ref{eq44}) and (\ref{eq44}) gives
\be\label{eq50}
f(R)=\frac{1}{2}\left(2\alpha-3\beta\right)R
\ee
which (normalizing the coefficient of $R$ to $1$) is the typical result of power-law expansion for the EH Lagrangian with pressureless dark matter content. In the other extreme case, when $B=0$ or equivalently in the far future at $t\rightarrow \infty$, one has $y\rightarrow \sqrt{A}=const.$, and all terms containing $dy/dx$ in (\ref{eq46}) disappear, leaving us with 
\be\label{eq51}
y\frac{df}{dy}+f(y)=6\alpha y
\ee
and for this equation one can write the general solution as
\be\label{eq52}
f(R)=3\alpha y+\frac{c}{y}=\frac{\alpha}{4}R+\frac{12c}{R}
\ee
where $c$ is the integration constant and in the last equality we used $y=\sqrt{A}$ and $R=12\sqrt{A}$ as follows from Eqs. (\ref{eq44}) and (\ref{eq45}) for $B=0$. By setting $c=0$ we obtain again the Einstein-Hilbert Lagrangian. An approximate solution to the Eq. (\ref{eq47}) can be obtained if we lower the power of the coefficients. For the four power of $y$ we will assume the following approximation
\be\label{eq53}
y^4=\left(A+B e^{-6x}\right)^2\simeq A\left(A+2B e^{-6x}\right)=2Ay^2-A^2
\ee
where we have neglected the term $B^2 e^{-12x}$, which after the appropriate normalization of $H$ at the present, contributes about $B^2\sim 0.09$ compared to the remaining terms in Eq. (\ref{eq53}) (at $x=0$) which contribute $\sim 0.91$. It should be noted that this approximation is valid for the present (or at low redshift) and becomes more accurate at future universe when $x$ becomes positive. By substituting $y^4$ from (\ref{eq53}) in the expressions between parentheses in (\ref{eq47}), then these expressions become of second order in $y$ and the powers of all the polynomial coefficients in eq. (\ref{eq47}) reduce by 2, giving
\be\label{eq54}
\begin{aligned}
&12y^3\left(Ay^2-A^2\right)\frac{d^2f(y)}{dy^2} -10y^2\left(Ay^2+A^2\right)\frac{df(y)}{dy}- 4y\left(Ay^2-2A^2\right)f(y)\\& =
-12\left(Ay^2-2A^2\right)\left[\left(2\alpha-3\beta\right)y^2+3\beta A\right]
\end{aligned}
\ee
The particular solution to this equation can be found as 
\be\label{eq55}
f(y)=\frac{\lambda_1}{y}+\lambda_2 y.
\ee
with
\be\nonumber
\lambda_1=\frac{5\alpha A}{2},\;\; \lambda_2=\frac{9\alpha}{4},\;\; \beta=-\frac{5\alpha}{24}
\ee
On the other hand, solving (\ref{eq45}) with respect to $y$ and replacing in (\ref{eq55}), leads to 
\be\label{eq56}
f(R)=\frac{\lambda_1}{\frac{R}{6}\pm \sqrt{\left(\frac{R}{6}\right)^2-3A}}+\lambda_2\left(\frac{R}{6}\pm \sqrt{\left(\frac{R}{6}\right)^2-3A}\right)
\ee
This solution was achieved at the expense of the above relation between $\alpha$ and $\beta$. The linear in $R$ term gives the EH Lagrangian by setting $\alpha=8/3$ which defines $\beta$ and $\lambda_1$ as $\beta=-5/9$ and $\lambda_1=20A/3$. Note that, provided $A\ne 0$ this function never reaches singularities and the expression under the square root is always positive, since the minimum value of $R$ as function of $y$, as follows from (\ref{eq45}) is $R_{min}=6\sqrt{3A}$, which gives the minimum value of the expression under the square root as $\left(\frac{R_{min}}{6}\right)^2-3A=0$. Its worth noting that although this model was obtained from a particular cosmological solution, namely the Chaplygin gas cosmology under the approximation (\ref{eq53}), nevertheless the model could be interesting by itself and deserves consideration as an $f(R)$ model independently of this solution, among other reasons because the linear in $R$ limit appears naturally. Both terms in (\ref{eq56}) could be relevant for both, late time cosmology and early universe (for large curvature with $R>>\sqrt{A}$, the first term in  (\ref{eq56}) becomes relevant, since the denominator can be very small if we assume the minus sign). Note also that one can obtain four different models, depending on the sign chosen for the two square roots in (\ref{eq56}), namely $(+ +)$, $(- -)$, $(+-)$ and $(-+)$. But applying the no-ghosts condition, $f_{,R}>0$, and stability (absence of tachyonic instability or consistence with matter-dominated epoch), $f_{,RR}>0$, the only combination that supports both conditions corresponds to $(- -)$. So, 
\be\label{eq57}
f(R)=R-\sqrt{R^2-108A}+\frac{40A}{R-\sqrt{R^2-108A}}
\ee
is the viable model that supports the conditions  $f_{,R}>0$ and  $f_{,RR}>0$. From these conditions follows that $R>R_0$, where $R_0=\frac{37\sqrt{A}}{\sqrt{10}}$ which is comparable with the $R_{min}=6\sqrt{3A}$. In fact the solution (\ref{eq56}) suggests the model
\be\label{eq58}
f(R)=R\pm\lambda_1\sqrt{R^2+\Lambda_1}+\frac{\lambda_2}{R\pm\lambda_3\sqrt{R^2+\Lambda_2}}
\ee
Assuming for instance, $\lambda_3=1, \lambda_1>0$ and $\Lambda_1=\Lambda_2=\Lambda>0$, the models obtained for the four possible combinations of signs, satisfy the conditions $f_{,R}>0$ and $f_{,RR}>0$. The following are the conditions for each case: \\
{\bf $(+ +)$}:
\be\nonumber
\lambda_2\le 0\;\;\&\;\; \lambda_1>-\lambda_2/\Lambda,\;\; or,\;\; \lambda_2>0,\;\;\&\;\;\Lambda\ge\lambda_2.
\ee
{\bf $(- -)$}:
\be\nonumber
\lambda_2<0,\;\; \&\;\; 0<\lambda_1<-\frac{\lambda_2}{\Lambda}.
\ee
{\bf $(+-)$}:
\be\nonumber
\begin{aligned}
&\lambda_2\le 0,\;\; {\it or}\;\; \lambda_2>0\;\;\&\;\; \Lambda=\lambda_2\;\; \&\;\; \lambda_1>\frac{2\lambda_2-\Lambda}{\Lambda}\\ &{\it or}\;\; \lambda_2>0\;\;\&\;\;\Lambda>\lambda_2\;\; \&\;\; \lambda_1>\frac{\lambda_2}{\Lambda}
\end{aligned}
\ee
{\bf $(-+)$}
\be\nonumber
\begin{aligned}
&\lambda_2>0\;\;\&\;\; \Lambda=\lambda_2;\;\&\;\;0<\lambda_1<1,\;\; or,\;\; \\ &\lambda_2>0\;\;\&\;\; \Lambda>\lambda_2\;\;\&\;\; 0<\lambda_1<\frac{\lambda_2}{\Lambda}
\end{aligned}
\ee
So, under the appropriate restrictions on $\lambda_1$, $\lambda_2$ and $\Lambda$ all the above models satisfy the restrictions for absence of ghosts and tachyonic degrees of freedom. 
\section{Discussion}
The late-time cosmological evolution of the universe presents unsolved challenges, which could demand approaches that seems consistent, at least at cosmological scales, where the asymptotical behavior of some gravity theories could contain a clue to the nature of the dark energy. 
In absence of a consistent theory of quantum gravity, the introduction of the holographic dark energy (which sets an infrared cutoff), can also be considered as the IR limit of the modified gravity, which is not inconsistent since the modified gravity is tested at local scales, while the HDE acts at cosmological scales. So, concerning the dark energy problem, and despite the fact that the $f(R)$ theories are sufficiently general and rich in solutions, connecting its large scale behavior with the HDE could render interesting results. Nevertheless, the above extension of HDE in modified gravity could rise theoretical questions, that deserve further studies.\\
In this paper we have presented a reconstruction of $f(R)$ gravity in the framework of the holographic dark energy, based on the holographic principle. A correspondence has been established between the effective density, coming from the modified gravity, and the holographic density, where the infrared cut-off proposed in \cite{granda1} was used to define the holographic density. 
The source term, which gives a non-homogeneous character to the modified Friedmann equation, allows to obtain explicit reconstructions for some relevant cosmic histories, with elementary functions of $R$. For the power law expansion, including the phantom solution, we have found that the particular solution for the non-homogeneous $f(R)$-equation, is the linear in $R$ term of general relativity. 
For a Little Rip solution we obtained a Lagrangian with Starobinsky, EH and cosmological constant terms, and it was shown that the reconstructed $f(R)$ can be reduced to the EH plus a cosmological constant term. Thus, in the frame of the HDE, the standard model consisting of EH with cosmological constant, can reproduce Little Rip solutions.   The $\Lambda$CDM-type solution was also considered and it was found that the reconstructed Lagrangian (the particular solution in the R-space) for $\Lambda$CDM-type cosmologies is the EH Lagrangian with cosmological constant, and this solution was possible, as in the power law case, due to the non homogeneous term in the equation (\ref{eq18}) represented by the holographic density. While in \cite{dunsby}, it was shown that to reconstruct the $\Lambda$CDM expansion requires the introduction of matter sources, in the present work this solution was obtained without introducing matter terms.\\
One of the problems of the method, used for the reconstruction of some important cosmic histories, is the fact that the obtained $f(R)$ models are given by very complicated analytical expressions, and its complexity make them of little use for further analysis. In the reconstruction of the Chaplygin gas cosmology presented here, it was not possible to find an exact analytical solution, but however, under a reasonable approximation as shown in Eq. (\ref{eq53}), it was found that the particular solution is given by simple functions of $R$ as seen from (\ref{eq56}), (\ref{eq57}). An important fact of this solution is that, as in the previous cases, the general relativity term emerges naturally from the reconstruction and the model is free of ghosts and tachyonic modes. In fact one can consider a family of similar models as proposed in (\ref{eq58}) which pass the test of no-ghosts and stability. 
 Both solution (\ref{eq57}) and its modification given by (\ref{eq58}), are interesting by themselves and deserve further analysis, since they are related with the Chaplygin gas cosmology that unifies the early time matter dominance with late accelerated expansion. An important fact of the reconstructed $f(R)$ models, for the cosmic histories considered here, is that all the reconstructed Lagrangians contain the EH term, maintaining the general relativity limit. These few examples reveal the potential and richness of the modified gravity in its connection with some facts of the holographic principle (that trace back to quantum gravity effects), that can be applied to the reconstruction of relevant cosmic histories. 
\section*{Acknowledgments}
\noindent This work was supported by Universidad del Valle under project CI 71074 and by
COLCIENCIAS grant number 110671250405.

\end{document}